\documentclass[aip,pop,graphicx,reprint]{revtex4-1}

\usepackage{amsmath}
\usepackage{graphicx}
\usepackage{amssymb}
\usepackage{bm}
\usepackage{braket}
\usepackage{hyperref}

\begin{document}

\title{Magnetohydrodynamical equilibria with current singularities and continuous rotational transform}

\author{Yao Zhou}
\email{yaozhou@princeton.edu}
\affiliation{Princeton Plasma Physics Laboratory, Princeton, NJ 08543, USA}

\author{Yi-Min Huang}
\affiliation{Department of Astrophysical Sciences, Princeton University, Princeton, NJ 08544, USA}

\author{A. H. Reiman}
\affiliation{Princeton Plasma Physics Laboratory, Princeton, NJ 08543, USA}

\author{Hong Qin}
\affiliation{Princeton Plasma Physics Laboratory, Princeton, NJ 08543, USA}
\affiliation{School of Physical Sciences, University of Science and Technology of China, Hefei, Anhui 230026, China}

\author{A. Bhattacharjee}
\affiliation{Princeton Plasma Physics Laboratory, Princeton, NJ 08543, USA}
\affiliation{Department of Astrophysical Sciences, Princeton University, Princeton, NJ 08544, USA}
\affiliation{Center for Computational Astrophysics, The Flatiron Institute, 162 Fifth Avenue, New York, NY 10010, USA }

\date{\today}

\begin{abstract}
We revisit the Hahm--Kulsrud--Taylor (HKT) problem, a classic prototype problem for studying resonant magnetic perturbations and 3D magnetohydrodynamical equilibria. We employ the boundary-layer techniques developed by Rosenbluth, Dagazian, and Rutherford (RDR) for the internal $m=1$ kink instability, while addressing the subtle difference in the matching procedure for the HKT problem. Pedagogically, the essence of RDR's approach becomes more transparent in the reduced slab geometry of the HKT problem.
We then compare the boundary-layer solution, which yields a current singularity at the resonant surface, to the numerical solution obtained using a flux-preserving Grad--Shafranov solver. The remarkable agreement between the solutions demonstrates the validity and universality of RDR's approach. In addition, we show that RDR's approach consistently preserves the rotational transform, which hence stays continuous, contrary to a recent claim that RDR's solution contains a discontinuity in the rotational transform.

\end{abstract}

\maketitle

\section{Introduction}
Ideal magnetohydrodynamics (MHD), a fundamental model in plasma physics, allows for the existence of tangentially discontinuous magnetic fields, i.e., current singularities. These singularities not only are mathematically intriguing, but also can have profound practical consequences. \citet{Grad1967} first proposed that smooth 3D MHD equilibria with nested toroidal flux surfaces generally do not exist, due to the pathologies that arise at rational surfaces. This theory has greatly impacted the studies of intrinsically non-axisymmetric magnetic fusion devices such as stellarators, as well as nominally axisymmetric ones like tokamaks, since they can be subject to resonant magnetic perturbations (RMPs) \citep{Garabedian1998,Boozer2005,Helander2014}. 

Rosenbluth, Dagazian, and Rutherford (RDR) \cite{Rosenbluth1973} first demonstrated how current singularities can dynamically emerge at resonant surfaces. They studied the internal $m=1$ kink instability in the cylindrical tokamak, and obtained a nonlinear boundary-layer solution to the ideal perturbed equilibrium, which contains a current singularity. It has since been realized that their approach is a powerful tool that can be applied to RMPs in general, whether spontaneous or forced \cite{Boozer2010,Loizu2017}. A resistive treatment of the $m=1$ kink-tearing mode has also been developed based on this approach\cite{Waelbroeck1989}.

RDR's classic paper invoked some subtle approximations that can be confusing at times. For instance, as discussed in Appendix\,\ref{oversight}, a seeming inconsistency in RDR's variable definitions has been misinterpreted, which eventually contributed to the recent claim that RDR's solution contains a discontinuity in the rotational transform\cite{Loizu2017}. We clear up this misinterpretation by demonstrating that RDR's solution consistently preserves the magnetic flux, within the approximations that are invoked. It follows that the rotational transform remains invariant and continuous, even though the magnetic field becomes discontinuous. 

Therefore, from a pedagogical perspective, we feel that it is worthwhile to present the fundamentals of RDR's approach in a more transparent manner, freeing it from the approximations that are secondary. With this mindset, we consider the reduced slab model of RMPs, following Boozer and Pomphrey (BP) \cite{Boozer2010}. However, BP did not construct a boundary-layer solution by matching the inner-layer solution to an outer-region solution, and that is what we shall do in this paper. 

Specifically, we apply RDR's boundary-layer approach to the ideal Hahm--Kulsrud--Taylor (HKT) problem \cite{Hahm1985}, which is arguably the simplest prototype problem for studying RMPs in slab geometry. We address the subtle difference in the matching procedure, since the RMP here is induced by external forcing. The contribution from matching the $1/x$ term turns out to be a second-order correction, which, however, needs to be included. Our work serves as a demonstration of the universality of RDR's boundary-layer approach, by rigorously applying it to an externally applied RMP.

We compare our boundary-layer solution with the numerical solution obtained using a flux-preserving Grad--Shafranov (GS) solver, which previously confirmed the formation of current singularity in the ideal HKT problem \cite{Zhou2016}. The solutions agree well, especially when the second-order correction from matching the $1/x$ term is included. Therefore, our work is also a direct quantitative validation of RDR's approach against numerical solution. (A previous numerical validation of RDR's results \cite{Park1980} is indirect, by contrast.)

This paper is organized as follows. In Sec.\,\ref{HKT}, the setup of the ideal HKT problem is introduced. In Sec.\,\ref{solution}, we derive the boundary-layer solution, namely, the asymptotically matched inner layer and outer-region solutions. The invariance of the rotational transform is also shown. In Sec.\,\ref{numerical}, we compare the boundary-layer solution to the numerical solution. Finally, our results are summarized and discussed in Sec.\,\ref{summary}.

\section{The ideal HKT problem}\label{HKT}
The HKT problem \cite{Hahm1985} considers, in slab geometry, an incompressible plasma magnetized by a sheared field. In order to relate to the original notations of RDR \cite{Rosenbluth1973}, we denote the coordinates as $(x,\phi,\zeta)$, where $x$ is the ``radial'' position (relative to the resonant surface), {and $\phi$ and $\zeta$ are the slab equivalents of, strictly speaking, the helical angles in RDR's cylindrical geometry}. For clarity, we term $\phi$ as the ``tangential'' direction, and $\zeta$ as ``helical''.
The system is periodic in $\phi$ and $\zeta$, with periods of $2\pi s$ and $2\pi R$, respectively. Here, we introduce the effective minor radius (of the resonant surface) $s$ and major radius $R$, which are both constants. 

Initially, the in-plane magnetic field reads $B_{0\phi}=j_0x$, which reverses sign at the resonant surface, $x=0$. The flux function can be written as $\psi_0(x)=j_0x^2/2$, where $j_0$ is a constant denoting the initial current density. The guide field $B_{0\zeta}$ is a large constant. This initial setup is a reduced MHD equilibrium. It coincides with the inner-layer approximation by RDR \cite{Rosenbluth1973}, and serves as a general slab model of the resonant layers of RMPs (with large aspect ratio $R/s$)\cite{Boozer2010}.

In the HKT problem, the flux surfaces at $x=\pm a$ are subject to mirrored sinusoidal perturbations, expressed in terms of the radial displacement, $\xi(\pm a,\phi)=\mp\delta\cos\phi$. It was originally proposed to study the forced magnetic reconnection induced by RMPs. Here, we only focus on the ``ideal stage'', assuming that the plasma is perfectly conducting, and hence the magnetic topology is invariant.
Suppose that the plasma relaxes to a new equilibrium state, which is a GS equilibrium and satisfies the 2D reduced MHD equilibrium equation,
\begin{equation}
\nabla^2\psi =\left(\partial_r^2+s^{-2}\partial_\phi^2\right)\psi= j_\zeta(\psi), \label{equilibrium}
\end{equation}
since the system is symmetric in $\zeta$. Here $\psi(r,\phi)$ is the perturbed flux function in Eulerian labeling $(r,\phi)$. Note that we use $r$ to denote the perturbed ``radial'' position in our slab geometry.

Given the HKT boundary conditions, multiple solutions to Eq.\,\eqref{equilibrium} exist \cite{Dewar2013}, but not all of them preserve the initial magnetic topology. For clarity, we refer to finding a topology-preserving equilibrium solution as the ideal HKT problem in this paper. 
The key to this problem is to construct the solution in terms of the radial mapping of the flux surfaces in hybrid Lagrangian-Eulerian labeling, $r(x,\phi)$ [or equivalently, $\xi(x,\phi)\equiv r-x$]. Tangential ($\phi$) flux conservation is then guaranteed by
\begin{equation}
\psi[r(x,\phi),\phi]=\psi_0(x).\label{psi}
\end{equation}
Here $x$ is the unperturbed radial position, or Lagrangian labeling, of the flux surface. It can also be considered as a flux coordinate. Meanwhile, helical ($\zeta$) flux conservation is guaranteed by the incompressibility constraint,
\begin{equation}
\langle\xi(x,\phi)\rangle\equiv \int^{2\pi}_0\frac{\mathrm{d}\phi}{2\pi} \,\xi(x,\phi)=0.\label{incomxi}
\end{equation}
Here the angle bracket denotes averaging over $\phi$ on a flux surface. Physically, Eq.\,\eqref{incomxi} means that the displacement $\xi$ on a flux surface $x$ does not change the volume (area) it encloses. Equations \eqref{equilibrium}-\eqref{incomxi}, together with the initial and boundary conditions, determine the solution to the ideal HKT problem.

\section{Boundary-layer solution}\label{solution}
{The linear solution \cite{Hahm1985,Zweibel1987} to the ideal HKT problem is known to introduce the so-called residual islands \cite{Boozer2010,Dewar2013} with widths of $O(\delta)$ on both sides of the resonant surface, which violate the preservation of magnetic topology. A proper boundary-layer treatment is thus needed in the vicinity of the resonant surface. 
In this section, we apply the boundary-layer approach developed by RDR \cite{Rosenbluth1973} to the ideal HKT problem.} In Secs.\,\ref{outer} and \ref{inner}, we present the outer region and inner-layer solutions, respectively. In Sec.\,\ref{iota}, we show that the rotational transform of the inner-layer solution consistently stays invariant and continuous. In Sec.\,\ref{matching}, we asymptotically match the inner layer and outer-region solutions to construct the boundary-layer solution.

\subsection{Outer-region solution}\label{outer}
When $|x|\gg\delta$, we can assume $|\partial_x\xi|\ll 1$, and linearize the system in terms of $\xi$. Specifically, we linearize $\nabla^2\psi $ using Eq.\,\eqref{psi}, such that
\begin{equation}
\nabla^2\psi =\psi_0'' +\psi_0'''\xi-\left(\partial_x^2+s^{-2}\partial_\phi^2\right)(\psi_0'\xi)+O(\xi^2). \label{linearN}
\end{equation}
From the boundary conditions we know that $\xi\sim\cos\phi$, which is automatically incompressible. We then deduce that the linear solution to the equilibrium equation \eqref{equilibrium} must satisfy
\begin{equation}
\left(\partial_x^2+s^{-2}\partial_\phi^2\right)(\psi_0'\xi)=\psi_0'''\xi. \label{linearE}
\end{equation}
Using the initial condition $\psi_0=j_0x^2/2$, we can obtain the linear solution to the ideal HKT problem \cite{Zweibel1987}, which contains two branches:
 \begin{equation}
\xi(x,\phi) = x^{-1}[C\sinh(|x|/s)+D\cosh(x/s)]\cos\phi\label{linear}.
\end{equation}
Note that we force $\xi$ to be odd in $x$, due to the parity of the ideal HKT problem.
Here $C$ and $D$ are constant coefficients to be determined. The boundary conditions at $x=\pm a$ give one constraint,
 \begin{equation}
C\sinh(a/s)+D\cosh(a/s)=-\delta a.\label{bc}
\end{equation}
Meanwhile, let us examine the asymptote of the linear solution \eqref{linear} approaching the resonant surface, $x=0$,
\begin{equation}
\xi(x\rightarrow 0^{\pm},\phi) = (\pm C/s+D/x)\cos\phi.\label{mathingout}
\end{equation}
Since the $1/x$ term diverges at $x=0$, $D$ is often taken to be zero, so that we have $C=-\delta a/\sinh(a/s)$ from Eq.\,\eqref{bc}. 

However, even with $D=0$, the linear displacement \eqref{linear} is discontinuous at $x=0$. This discontinuity results in a jump in the perturbed tangential magnetic field \cite{Hahm1985}, 
\begin{equation}
\delta B_\phi =\text{sgn}(x) j_0a\delta \cosh (x/s) \cos \phi/ [s\sinh (a/s)],
\end{equation}
i.e., a current singularity. Here $\text{sgn}(x)$ denotes the sign function. {Notably, this discontinuity also introduces the aforementioned residual islands by inducing the overlapping of flux surfaces, which is not physically permissible.}

In the vicinity of the resonant surface ($|x|\ll\delta$), the linear assumption $|\partial_x\xi|\ll 1$ breaks down, and naturally the linear solution becomes invalid. The proper treatment of this resonant layer requires a boundary-layer procedure, which entails constructing an inner-layer solution and matching it asymptotically to the outer-region solution \eqref{linear}, using the matching condition given by Eq.\,\eqref{mathingout}. Note that we do not discard the $1/x$ term here, and its role will be explained in Sec.\,\ref{matching}.

As BP \cite{Boozer2010} discussed, the inner-layer solution that RDR \cite{Rosenbluth1973} derived for the internal $m=1$ kink problem can readily be applied here, without modifications. Their derivation is replicated in Sec.\,\ref{inner}. 

\subsection{Inner-layer solution}\label{inner}
RDR argued that in the resonant layer ($|x|\ll\delta$), the tangential variation is much slower than radial. Hence, they neglected the $s^{-2}\partial_\phi^2\psi$ term in Eq.\,\eqref{equilibrium}, ending up with
\begin{equation}
\partial_r^2\psi = j_\zeta(\psi). \label{equilibriumS}
\end{equation}
In a genuine slab system, such as the HKT problem that we are considering, this is the only approximation needed in the inner layer. Notably, this approximation does not affect flux conservation.

Integrating Eq.\,\eqref{equilibriumS} leads to a general solution,
\begin{equation}
 (\partial_r\psi)^2 = j_0^{2}[F(\psi)+g(\phi)],\label{FG}
\end{equation}
where $F$ and $g$ are arbitrary functions. Then, we can use the invariance of the flux function \eqref{psi} to rewrite Eq.\,\eqref{FG} in terms of the radial mapping $r(x,\phi)$,
\begin{equation}
\partial_x r = {|x|}/{\sqrt{f(x)+g(\phi)}},\label{RX}
\end{equation}
with $f(x)=F[\psi_0(x)]=F(j_0{x}^2/2)$. The absolute value is introduced by the monotonicity condition, $\partial_x r\ge 0$, so that the flux surfaces stay nested.
With $\partial_x \xi=\partial_x r-1$, we can integrate Eq.\,\eqref{RX} and obtain the displacement, 
\begin{equation}
\xi(x,\phi) = h(\phi)+\int_0^x\mathrm{d}x'\left[\frac{|x'|}{\sqrt{f(x')+g(\phi)}}-1\right]\label{xis},
\end{equation}
with $h$ being another arbitrary function. 

Meanwhile, the tangential magnetic field is given by \cite{Boozer2010}
\begin{equation}
B_\phi(x,\phi) = \partial_r\psi = \text{sgn}(x)j_0\sqrt{f(x)+g(\phi)}\label{Bs}.
\end{equation}
The finite jump at $x=0$ corresponds to a delta-function (surface) current, 
\begin{equation}
I'_\delta(\phi) = 2j_0\sqrt{f(0)+g(\phi)}\label{Is}.
\end{equation}
Loizu and {Helander} (LH) \cite{Loizu2017} termed this current singularity as ``zonal'', since its flux-surface average $\langle I'_\delta \rangle$ is non-zero. (This does not imply banded radial structure, however.) Interestingly, as we will show next, $I'_\delta $ itself must be zero somewhere \cite{Boozer2010}.

The solution is also subject to the incompressibility constraint \eqref{incomxi}.
Differentiating with respect to $x$, we have
\begin{equation}
\langle \partial_x r\rangle=1\label{incom}.
\end{equation}
Now, using Eq.\,\eqref{RX}, we obtain
\begin{equation}
 \left\langle\left[{f(x)+g(\phi)}\right]^{-1/2}\right\rangle=|x|^{-1}\label{incoms}.
\end{equation}
The fact that the right-hand side diverges at $x=0$ suggests that $f(0)+g(\phi)$ (and accordingly, $I'_\delta$) must be zero somewhere. That is, $f(0)=-g_{\text{min}}$. With the freedom to choose $f(0)$, we set $f(0)=-g_{\text{min}}=0$ for convenience.

So far, we have derived the general form of RDR's inner-layer solution, Eqs.\,\eqref{xis} and \eqref{incoms}, in terms of functions $f$, $g$, and $h$. These functions are further determined by matching the solution \eqref{xis} to the outer-region solution to the specific problem. However, before carrying out such a procedure for the ideal HKT problem in Sec.\,\ref{matching}, we take a detour in Sec.\,\ref{iota} and show that RDR's inner-layer solution, in its general form, consistently preserves the rotational transform, which hence stays continuous as initially prescribed.

\subsection{Rotational transform}\label{iota}
Recall that the rotational transform can be expressed in terms of the flux functions, and hence should stay invariant as long as the magnetic flux is preserved. Here, we explicitly demonstrate such invariance in our slab system.
Consider the field line flow,
\begin{equation}
\frac{s\mathrm{d}\phi}{B_{\phi}}=\frac{R\mathrm{d}\zeta}{B_{\zeta}},
\end{equation}
where $s$, $R$ and $B_{\zeta}$ are all constants, and the system is symmetric in $\zeta$. Then, we can integrate over one period in $\phi$, and calculate the increment in $\zeta$,
\begin{equation}
\Delta\zeta=\frac{sB_{\zeta}}{R}\int_0^{2\pi}\frac{\mathrm{d}\phi}{B_{\phi}}.
\end{equation}
The rotational transform in our slab system then reads
\begin{equation}
\iota=\frac{2\pi}{\Delta\zeta}=\frac{R}{sB_{\zeta}}\left\langle B_{\phi}^{-1}\right\rangle^{-1}.\label{iotaslab}
\end{equation}

Now, let us consider flux conservation. Tangential ($\phi$) flux conservation is guaranteed if the tangential magnetic field is derived from the flux function, Eq.\,\eqref{psi}:
\begin{equation}
B_\phi= \partial_r\psi=\psi_0'/\partial_{x}r=B_{0\phi}/\partial_{x}r.\label{tslab}
\end{equation}
Here $\mathbf{B}_0$ denotes the initial magnetic field that does not depend on $\phi$. Substituting into Eq.\,\eqref{iotaslab}, we have
\begin{equation}
\iota=\frac{R}{sB_{\zeta}}\left\langle B_{\phi}^{-1}\right\rangle^{-1}=\frac{RB_{0\phi}}{sB_{\zeta}}\left\langle{\partial_{x}r}\right\rangle^{-1}.\label{iotaslabs}
\end{equation}

{Meanwhile, helical ($\zeta$) flux conservation corresponds to the incompressibility constraint \eqref{incom}, and $B_{\zeta}=B_{0\zeta}$. Combining with Eq.\,\eqref{iotaslabs}, we have, as expected,} 
\begin{equation}
\iota=\frac{R{B_{0\phi}}}{sB_{\zeta}}=\frac{R{B_{0\phi}}}{sB_{0\zeta}}.\label{iota0}
\end{equation}
That is, $\iota$ stays invariant as initially prescribed in our slab system.

So far, we have shown that flux conservation guarantees the invariance of the rotational transform, and not discussed RDR's inner-layer solution specifically. Still, it should be obvious by now that RDR's solution, which conserves the magnetic flux [Eqs.\,\eqref{incom} and \eqref{tslab}], should automatically preserve the rotational transform $\iota$. Next, we explicitly calculate $\iota$ in RDR's solution, and show that it indeed consistently stays invariant and continuous.

Using RDR's solution \eqref{Bs}, which satisfies Eq.\,\eqref{tslab} and preserves the tangential flux, Eq.\,\eqref{iotaslab} becomes
\begin{equation}
 \iota=\frac{R}{sB_{\zeta}}\left\langle B_{\phi}^{-1}\right\rangle^{-1}=
\frac{ \text{sgn}(x)j_0R}{sB_{\zeta}\left<{{[f(x)+g(\phi)]}}^{-1/2}\right>}.
\end{equation}
Then, using the incompressibility constraint \eqref{incoms}, which follows from Eq.\,\eqref{incom} and preserves the helical flux, we have
 \begin{equation}
\iota=\frac{ Rj_0x}{sB_{\zeta}}=\frac{ RB_{0\phi}}{sB_{0\zeta}}.
\end{equation}
Note that $B_{0\phi}=j_0x$ and $B_{\zeta}=B_{0\zeta}$. As expected, $\iota$ stays invariant and continuous as initially prescribed, regardless of the exact forms of $f$ and $g$. The point is that RDR's solution is constructed in terms of the incompressible displacement of the flux surfaces, which automatically guarantees flux conservation in our system.

At first glance, the fact that the tangential magnetic field is discontinuous whereas the rotational transform is continuous may seem somewhat counter-intuitive. The key subtlety here that makes this possible is that $B_\phi(0^{\pm},\phi)$ has to be zero somewhere, so that $\langle B_\phi^{-1}\rangle$ diverges, and $\iota$ can stay well-behaved at $x=0$.

The recent claim by LH\cite{Loizu2017}, that RDR's solution contains a discontinuity in the rotational transform, originates from a misinterpretation of RDR's slab approximation of the resonant layer (see Appendix\,\ref{oversight}). By considering a genuine slab system here, we see that the issue resides not in RDR's approximations, but LH's calculation of the rotational transform, seemingly from $\langle B_\phi\rangle$ rather than $\langle B_\phi^{-1}\rangle^{-1}$ [Eqs.\,(10) to (11) therein]. As shown above, the (properly calculated) rotational transform in RDR's inner-layer solution consistently stays invariant and continuous, even though the magnetic field is discontinuous.

\subsection{Matching}\label{matching}
Now, let us match RDR's general inner-layer solution, derived in Sec.\,\ref{inner}, to the outer-region solution to the ideal HKT problem given in Sec.\,\ref{outer}. 
First, we examine the asymptote of the inner-layer solution \eqref{xis} as $|x|\rightarrow\infty$. {Using the fact that $f(x)=F(j_0x^2/2)$ is an even function, we rewrite Eq.\,\eqref{xis} as}
\begin{align}
\xi(x,\phi) =& h(\phi)+\text{sgn}(x)\int_0^{|x|}\mathrm{d}x'\left[\frac{x'}{\sqrt{f(x')+g(\phi)}}-1\right]\label{xis1}
%\nonumber\\
%=&h(\phi)+\text{sgn}(x)\Bigg\{ \int_0^{\infty}\mathrm{d}x\left[\frac{x}{\sqrt{f(x)+g(\phi)}}-1\right]\nonumber\\
%&-\int_{|x|}^{\infty}\mathrm{d}x'\left[\frac{x'}{\sqrt{f(x')+g(\phi)}}-1\right]\Bigg\},
\end{align}
{Meanwhile, from Eq.\,\eqref{incoms}, we can infer that as $|x|\rightarrow\infty$, $f(x)\rightarrow x^2$, and hence}  
\begin{align}
\int_{|x|}^{\infty}\mathrm{d}x'\left[\frac{x'}{\sqrt{f(x')+g(\phi)}}-1\right]\rightarrow-\frac{g(\phi)}{2|x|}.
\end{align}
Then, the asymptote of the inner-layer solution reads
\begin{equation}
%\xi(x\rightarrow\pm\infty,\phi) = h(\phi)\pm\int_0^{\infty}\mathrm{d}x\left(\frac{x}{\sqrt{f+g}}-1\right)+\frac{g(\phi)}{2x}.\label{infty}
\xi(x\rightarrow\pm\infty,\phi) = h(\phi)\pm\int_0^{\infty}\mathrm{d}x\left[\frac{x}{\sqrt{f(x)+g(\phi)}}-1\right]+\frac{g(\phi)}{2x}.\label{infty}
\end{equation}

The matching conditions are obtained by comparing Eq.\,\eqref{infty} to the asymptote of the outer-region solution \eqref{mathingout}. Specifically, matching the constant term gives $h(\phi)=0$, as well as
\begin{equation}
\int_0^{\infty}\mathrm{d}x\left[\frac{x}{\sqrt{f(x)+g(\phi)}}-1\right]=(C/s)\cos\phi.
\end{equation}
RDR then used Eq.\,\eqref{incoms} to eliminate $x$, and obtained a fundamental integral equation for $g(\phi)$,
\begin{align}
\int_0^{\infty}&\mathrm{d}f\left\langle(f+g)^{-1/2}\right\rangle^{-3}\left\langle(f+g)^{-3/2}\right\rangle&\nonumber\\
&\times\left[(f+g)^{-1/2}-\left\langle(f+g)^{-1/2}\right\rangle\right]=(2C/s)\cos\phi.&\label{integral}
\end{align}

To actually solve Eq.\,\eqref{integral} is, quoting RDR, ``an almost impossible task''. RDR constructed a functional that is extremized by Eq.\,\eqref{integral}, which in turn can be solved variationally. LH \cite{Loizu2017} implemented this procedure  numerically and obtained a solution of $g(\phi)$. In principle, one could proceed using their numerical solution.

Meanwhile, RDR also found that a rough form for the solution would be $g\sim\cos^8(\phi/2)$. Moreover, LH showed that their numerical solution of $g(\phi)$ can be well approximated by
\begin{equation}
g(\phi)\approx 4C^2/(3s^2)\cos^8(\phi/2).\label{gapprox}
\end{equation}
In order to keep our derivation analytically tractable, we choose to proceed using this approximate form of $g(\phi)$. Readers interested in the details on how to solve for $g(\phi)$ are referred to Refs.\,\onlinecite{Rosenbluth1973,Loizu2017}. 

Now, let us consider the $1/x$ term. Obviously, $g(\phi)/2$ and $D\cos\phi$ cannot be matched exactly here. Following RDR, we expand $g(\phi)=\sum\Gamma_m\cos(m\phi)$, and only match the $m=1$ component,
\begin{equation}
D=\Gamma_1/2=\left\langle g(\phi)\cos\phi\right\rangle\approx7C^2/(24s^2).\label{Ds}
\end{equation}
Here the approximate form of $g(\phi)$ \eqref{gapprox} is used. Substituting Eq.\,\eqref{Ds} into Eq.\,\eqref{bc}, we have an equation that determines the coefficient $C$,
 \begin{equation}
7C^2\cosh(a/s)/(24s^2)+C\sinh(a/s)+\delta a=0.
\end{equation}
This equation has two roots. We keep only one of them,
 \begin{equation}
C=\frac{\sqrt{\sinh^2(a/s)-7\delta a\cosh(a/s)/(6s^2)}-\sinh(a/s)}{7\cosh(a/s)/(12s^2)},\label{Cs}
\end{equation}
as it correctly approaches the $C=-\delta a/\sinh(a/s)$ limit when $\delta\ll a$.

It is worthwhile to address the subtle difference between our matching procedure and RDR's for the internal $m=1$ kink problem. In the ideal HKT problem, the outer-region solution is driven by external forcing with known amplitude $\delta$. {The contribution from matching the $1/x$ term is of second order in $\delta$ [$D\sim O(\delta^2)$]}, and may seem negligible when $\delta$ is small. However, as we will show in Sec.\,\ref{numerical}, this second-order correction becomes visible when $\delta$ is relatively large.

In contrast, in the internal $m=1$ kink problem, the outer-region solution is driven by an instability with its amplitude to be determined. It is the ratio between the constant term and the $1/x$ term that is known. Roughly speaking, one may interpret it as if $C/D$ is given instead of Eq.\,\eqref{bc}, and then $C$ and $D$ can be determined using Eq.\,\eqref{Ds}. In this case, matching the $1/x$ term is critical in determining the amplitude of the displacement, and its contribution cannot be neglected. 

Finally, let us summarize the boundary-layer solution we have derived: the outer-region solution is Eq.\,\eqref{linear}, with the coefficients given by Eqs.\,\eqref{Ds} and \eqref{Cs}; the inner-layer solution is Eq.\,\eqref{xis}, with $g(\phi)$ given by Eq.\,\eqref{gapprox}, $f(x)$ subsequently determined (numerically) by Eq.\,\eqref{incoms}, and $h(\phi)=0$.
In Sec.\,\ref{numerical}, we compare this boundary-layer solution with the fully nonlinear solution to the ideal HKT problem obtained using a flux-preserving GS solver.
As will be shown, the agreement between the solutions is quite impressive, even though we are using the approximate form of $g(\phi)$ \eqref{gapprox}.

\section{Comparison with numerical solution}\label{numerical}
The ideal HKT problem has previously been studied numerically \cite{Zhou2016}, using two different methods: one is a fully Lagrangian, moving-mesh variational integrator for ideal MHD \cite{Zhou2014}; the other is a flux-preserving GS solver, where the equilibrium guide field is determined self-consistently to preserve its flux at each flux surface \cite{Huang2009}, unlike conventional ones where it is prescribed as a function of the flux. Both methods preserve the magnetic flux exactly; otherwise, they would not be capable of studying the ideal HKT problem.
The numerical solutions from the two methods agree, and confirm that the solution to the ideal HKT problem contains a current singularity \cite{Zhou2016}. The GS solver shows better convergence at the resonant surface, since it employs the same hybrid Lagrangian-Eulerian labeling as in this paper. Therefore, we use the GS solution as the benchmark for the boundary-layer solution. The fully Lagrangian variational integrator proves useful where the GS solver does not apply, such as problems with more complex magnetic topology \cite{Zhou2017b} or in 3D \cite{Zhou2017}.

For convenience, when solving the ideal HKT problem numerically, we do not enforce incompressibility. Instead, we initialize with a strong guide field $B_{0\zeta}=\sqrt{B_0^2-j_0^2x^2}$, such that the unperturbed equilibrium is force-free. This minor alteration in setup, due to Zweibel and Li \cite{Zweibel1987}, does not affect the physics of the ideal HKT problem, since the plasma is still close to incompressible, especially near the resonant surface. After all, here incompressibility itself is an approximation that is valid in the strong-guide-field limit.

The parameters that we use to obtain the numerical solution are $a=0.5$, $s=1/\pi$, $\delta=0.1$, $j_0=1$, and $B_0=10$. Given the parity of the ideal HKT problem, we only need to consider a half of the domain, $x\in[0^+,a]$, with boundary condition $\xi(0,\phi)=0$. The solution in the other half, $x\in[-a,0^-]$, is given by $\xi(x,\phi)=-\xi(-x,\phi)$.

\begin{figure}
\includegraphics[scale=0.45]{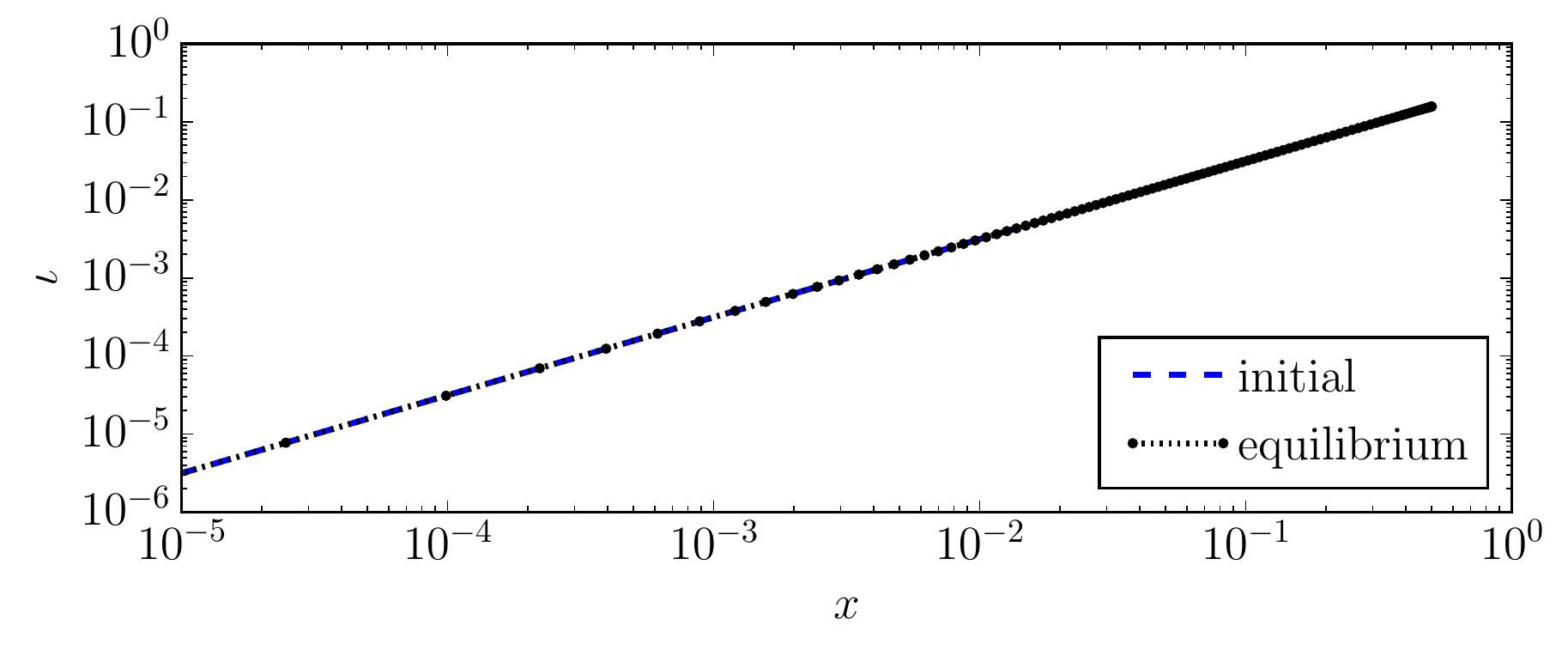}
\caption{\label{iotap}The rotational transform $\iota$ of the equilibrium solution (dotted) from the GS solver agrees perfectly with the initial profile (dashed). $R=1$ is used for plotting.}
\end{figure}

In Fig.\,\ref{iotap}, we show the rotational transform of the numerical solution calculated using Eq.\,\eqref{iotaslab}, which still applies here since $B_\zeta$ is constant on the flux surfaces in a GS equilibrium. The equilibrium solution agrees with the rotational transform profile of the initial state \eqref{iota0}, confirming its supposed preservation of the magnetic flux. Figure \ref{iotap} also serves as a sanity check for the conclusions in Sec.\,\ref{iota}, including the validity of Eq.\,\eqref{iotaslab} and that the rotational transform of the solution to the ideal HKT problem stays continuous, despite the discontinuous magnetic field.

\begin{figure}
\includegraphics[scale=0.45]{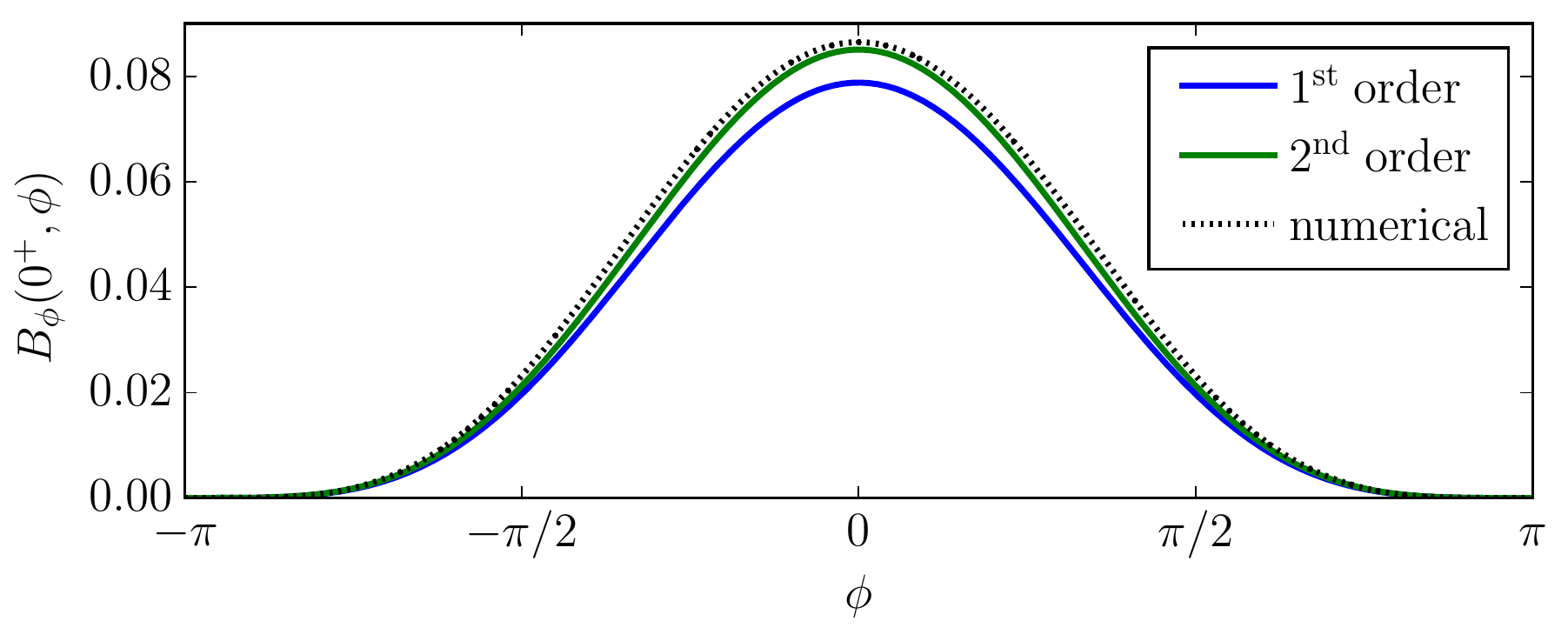}
\caption{\label{Byp}At $x=0^+$, the boundary-layer solutions (solid) of the tangential magnetic field $B_\phi$ are compared with the numerical solution (dotted). The solution from second-order matching (green) shows better agreement than that from first-order matching (blue).}
\end{figure}

Now, let us compare the boundary-layer solution with the numerical solution. In Fig.\,\ref{Byp}, we show the tangential magnetic field at the resonant surface, $B_\phi(0^+,\phi)$, of both the inner-layer solution \eqref{Bs} and the numerical solution. Figure \ref{Byp} is also an illustration of the structure of the current singularity, since the delta-function current on the resonant surface $I_\delta'(\phi)=2B_\phi(0^+,\phi)$. 

Here we refer to the boundary-layer solution obtained in Sec.\,\ref{matching}, with $C$ and $D$ given by Eqs.\,\eqref{Ds} and \eqref{Cs}, as second-order, since it accounts for the second-order correction from matching the $1/x$ term. In addition, we also show the first-order solution without matching the $1/x$ term, i.e., with $C=-\delta a/\sinh(a/s)$ and $D=0$. The second-order solution agrees with the numerical solution significantly better than the first-order solution, because the perturbation we use ($\delta=0.1$) is relatively large. In this sense, Fig.\,\ref{Byp} clearly demonstrates the existence and importance of the second-order correction.

\begin{figure}
\includegraphics[scale=0.45]{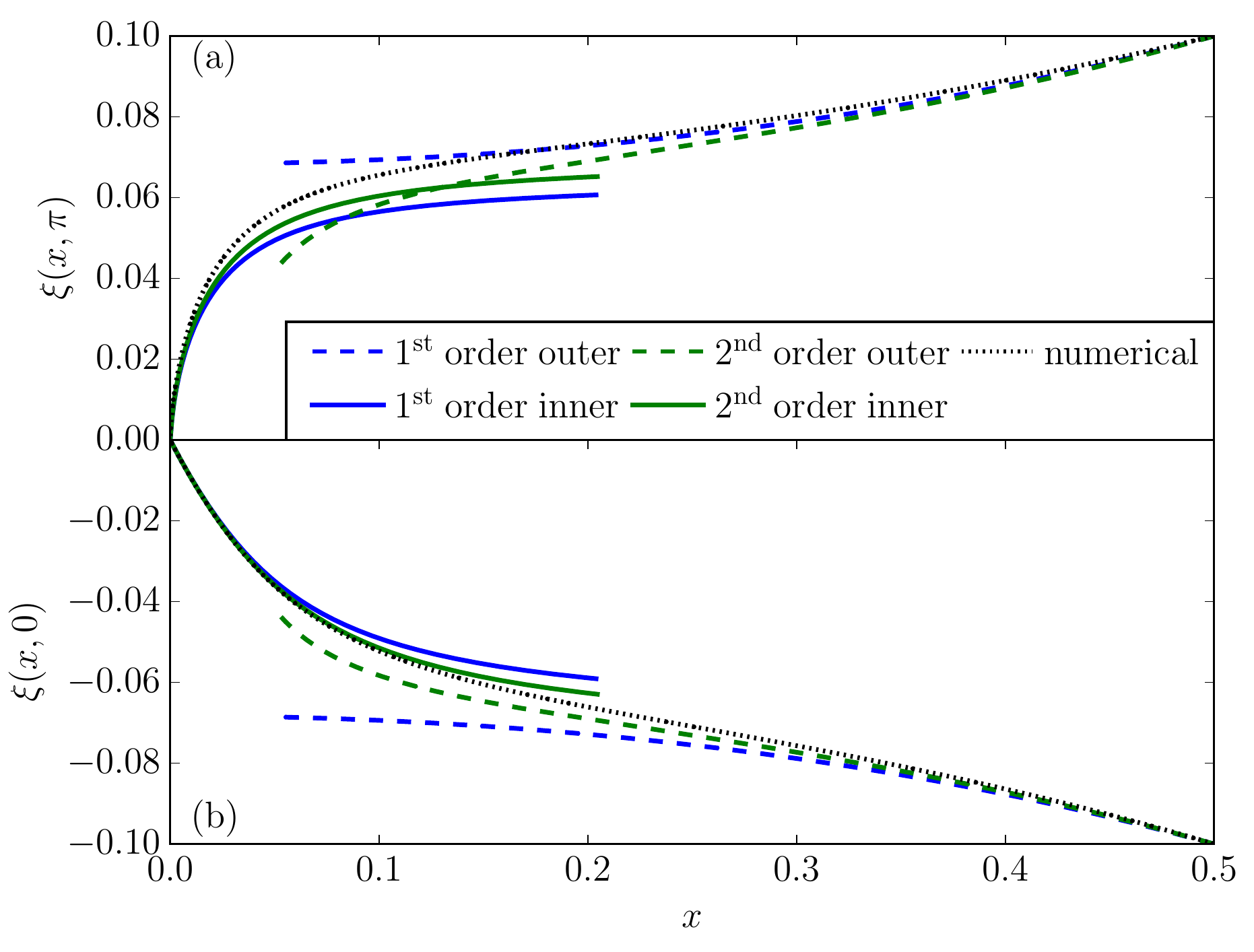}
\caption{\label{xip}At $\phi=\pi$ (a) and $\phi=0$ (b), the boundary-layer solutions of the radial displacement $\xi$, including inner layer (solid) and outer-region solutions (dashed), are compared with the numerical solutions (dotted). The solutions from second-order matching (green) exhibit better agreement than those from first-order matching (blue).}
\end{figure}

In Fig.\,\ref{xip}, we compare the radial displacement $\xi$ of the boundary-layer solutions with the numerical solution, at $\phi=\pi$, where the plasma is stretched the most, and at $\phi=0$, where the plasma is compressed the most. Again, the second-order solutions show better agreement with the numerical solution than first-order. In particular, the second-order correction from the $1/x$ term in the outer-region solution brings it closer toward matching with the inner layer and the numerical solutions.

One might notice that in Fig.\,\ref{xip}, the agreement between the boundary-layer solutions and the numerical solution is slightly worse at $\phi=\pi$ than at $\phi=0$. The reason could be that we are using an approximate form of $g(\phi)$ \eqref{gapprox}, which may not be as accurate at $\phi=\pi$. It is possible that the agreement can be even better, had we used a more accurate form of $g(\phi)$, such as LH's numerical solution to Eq.\,\eqref{integral} \cite{Loizu2017}.

\section{Summary and discussion}\label{summary}
In this paper, we derive a boundary-layer solution to the ideal HKT problem, using the techniques that RDR developed originally for the internal $m=1$ kink problem. The subtle difference in the matching procedure between the two problems, due to the different natures of the RMPs, is addressed. In addition, we show that the rotational transform in RDR's inner-layer solution consistently stays invariant and continuous, contrary to the recent claim that it contains a discontinuity.

Then, we compare our boundary-layer solution with the numerical solution obtained from a flux-preserving GS solver. The solutions agree especially well when the second-order correction from matching the $1/x$ term is included, and definitively confirm the existence of ideal MHD equilibria with (zonal) current singularities and continuous rotational transform. On one hand, our work is a direct quantitative validation of RDR's boundary-layer approach, the inner-layer solution in particular. On the other hand, it demonstrates the universality of RDR's approach regardless of the nature of the RMP, be it an instability as in the internal $m=1$ kink problem, or from external forcing as in the ideal HKT problem here. 

So far, we have intentionally restricted our discussions entirely within the slab geometry. Pedagogically, we are trying to illustrate the fundamental features of RDR's boundary-layer approach in a more comprehensible way. Still, this is a reduced prototype problem, and one may reasonably wonder whether our conclusions stand in more realistic geometries.

In the internal $m=1$ kink problem, RDR reduced the resonant layer to a slab model, where the equilibrium is solved only to $O(\epsilon^0)$ ($\epsilon\equiv s^2/R^2$). This is actually sufficient, and consistent with the ordering of their solution: with amplitude $\xi\sim O(\epsilon)$, force balance is needed only to $O(\epsilon^0)$ for the solution to be $O(\epsilon)$ accurate. The point is that we believe RDR's kinked equilibrium solution, which contains a current singularity and continuous rotational transform, is valid under the approximations invoked. We expect that applying RDR's approach to RMPs of other nature in non-slab geometries should lead to similar conclusions, provided consistent ordering. 

Of course, further numerical validation is warranted in non-slab geometries, possibly using the fully Lagrangian variational integrator for ideal MHD \cite{Zhou2014}. The GS solver is no longer applicable here, while Eulerian methods require \textit{ad hoc} treatments to avoid artificial reconnection, such as the artificial fields that effectively remove the resonance in Ref.\,\onlinecite{Park1980}. Another possibility is to enable the Stepped-Pressure Equilibrium Code (SPEC) \cite{Hudson2012}, which can handle 3D MHD equilibria with current singularities\cite{Loizu2015, Loizu2015b}, to produce those with continuous rotational transform.

Finally, it would be interesting to consider generalizing RDR's approach to 3D line-tied geometry, where Parker's conjecture of current singularity formation \cite{Parker1972} has been controversial for decades. Notably, a recent numerical study on the ideal HKT problem in 3D line-tied geometry \cite{Zhou2017} shows that the current density distribution becomes more localized as the system length increases, but is inconclusive on whether the solution becomes singular at a finite length. One major analytical challenge here is, due to the lack of resonance in the absence of closed field lines, the system cannot be straightforwardly reduced to 2D as in the case of RMPs.

\acknowledgments
We thank D.\,Pfefferl\'e for exposing an oversight in the manuscript; J.\,Loizu, P.\,Helander, and S.\,Hudson for helpful discussions; and the anonymous reviewer for thoughtful suggestions that helped us improve our manuscript.
This research was supported by the U.S.~Department of Energy under Contract No.~DE-AC02-09CH11466.
A.\,Bhattacharjee acknowledges support by a grant from the Simons Foundation/SFARI (560651, AB).

\appendix
\section{a misinterpretation of RDR's approximation}\label{oversight}
In RDR's original treatment of the internal $m=1$ kink instability \cite{Rosenbluth1973}, $r$ is denoted as the perturbed (Eulerian) radius of a flux surface, while $x$ is its initial (Lagrangian) radius \textit{relative to} $s$, which is the unperturbed radius of the resonant surface. However, RDR then defined the radial displacement as $\xi= r - x$ [Eq.~(22) therein]. This is, strictly speaking, inconsistent with the definitions above, which locate the resonant surface at $r=s$ (initially) and $x=0$. The proper definition should be
\begin{equation}
\xi = r -r_0 =r - (x+s),\label{xid}
\end{equation}
where $r_0=x+s$ denotes the unperturbed radius of the flux surface. 

Interestingly, such an inconsistency does not appear to have affected the validity of their results, since RDR approximated the resonant layer $r=s\pm O(\epsilon)$ as a slab. In particular, the incompressibility constraint originally reads
\begin{equation}
\left\langle r\partial_x r\right\rangle=\left\langle r_0\partial_xr_0\right\rangle=r_0,
\end{equation}
in cylindrical geometry. Now, using Eqs.\,\eqref{RX} and \eqref{xid}, we have
\begin{equation}
\left\langle \frac{(s+x+\xi)|x|}{\sqrt{f(x)+g(\phi)}}\right\rangle=s+x.\label{incomc}
\end{equation}
Here, RDR essentially approximated both $r=s+x+\xi$ and $r_0=s+x$ as $s$, which is consistent with the slab approximation of the resonant layer. The incompressibility constraint then reduces to its slab version, Eq.\,\eqref{incoms}.

However, the original oversight by RDR in writing $\xi=r-x$ has caused some confusion down the road. That is, Loizu et al.\,\cite{Loizu2016} [Eq.\,(A6) therein] omitted the constant $s$ in Eq.\,\eqref{incomc}, and subsequently stated that RDR ignored $\xi$ with respect to $x$ to obtain Eq.\,\eqref{incoms}, which arguably would not consistently preserve the magnetic flux. This is a misinterpretation of RDR's ordering in the resonant layer, which is merely a slab approximation. Eventually, this misinterpretation contributed to the claim that RDR's inner-layer solution contains a discontinuity in the rotational transform \cite{Loizu2017}.

\end{document}